\pdfoutput=1
\documentclass[twocolumn,final,letterpaper]{IEEEtran}
\usepackage[utf8]{inputenc}
\usepackage[T1]{fontenc}
\usepackage{fixltx2e}
\usepackage{graphicx}
\usepackage{longtable}
\usepackage{float}
\usepackage{wrapfig}
\usepackage{soul}
\usepackage{textcomp}
\usepackage{marvosym}
\usepackage{wasysym}
\usepackage{latexsym}
\usepackage{amssymb}
\usepackage{multirow}
\usepackage{setspace}
\IEEEoverridecommandlockouts

\title{Asynchronous Source Clock Frequency Recovery through Aperiodic Packet Streams}
\author{Kyeong Soo Kim, \IEEEmembership{Member, IEEE}\thanks{K. S. Kim is with the College of Engineering, Swansea University, Swansea, SA2 8PP, Wales United Kingdom (e-mail: k.s.kim@swansea.ac.uk).}}
\date{Time-stamp: <2013-04-25 11:42:43 Kyeong Soo (Joseph) Kim>}

\begin{document}

\maketitle

\begin{abstract}
We consider the most general case of source clock frequency recovery (SCFR) in
packet networks, i.e., asynchronous SCFR through aperiodic packet streams, where
there is neither a common reference clock nor any relation between packet
generation intervals and a source clock frequency. We formulate the problem of
asynchronous SCFR with timestamps as a linear case of regression through the
origin (RTO) and propose two schemes, one based on recursive least squares (RLS)
method and the other based on simple heuristics of cumulative ratio of
interarrival and interdeparture times, which provide better estimates of the
source clock frequency with faster convergence than conventional phase-locked
loop (PLL)-based schemes.
\end{abstract}


\section{Introduction}
\label{sec-1}

\IEEEPARstart{R}{eal-time} variable bit rate (VBR) services carried over
asynchronous packet networks impose serious challenges on the source clock
frequency recovery (SCFR) at a receiver. First, there is no common reference
clock available at a source and a receiver, with which the source clock
frequency can be encoded and transmitted to the receiver for its recovery there
as in the synchronous residual time stamp (SRTS) method
\cite{I.363.1:96}. Secondly, unlike constant bit rate (CBR) services, packet
generation intervals (i.e., interdeparture times) at the source are varying over
time and have no relation with the source clock frequency. These, together with
unpredictable packet jitter and frequent packet losses, make the asynchronous
SCFR quite a difficult task.

A periodic packet stream is a stream of packets generated and transmitted at
regular intervals (i.e., constant interdeparture times) at a source, which is
typically related with a source clock frequency, while an aperiodic packet
stream has no such regular interval. Note that CBR services can result in
aperiodic packet streams when the packet size is not fixed. On the other hand,
the components of VBR traffic can be periodic like the first packet of each
encoded video frame. So we use the term \emph{aperiodic packet streams} in order to
clarify the major focus of the asynchronous SCFR schemes studied in this
paper. Note that, when there is no periodic time structure in generated packet
streams (e.g., voice over IP (VoIP)), these services are more tolerable to the
playout delay. The SCFR, however, is still useful even in such a case because
the receiver can eliminate the distortion caused by the lengthening or
shortening of silence periods between talkspurts with a recovered source clock,
which is an inevitable side effect of adaptive receiver buffer algorithms
\cite{melvin02:_time_voip}.

Existing SCFR schemes extract the information on the source clock frequency from
packet interarrival times (e.g., \cite{Kim:00-1,aweya06:_clock,santos12}),
buffer levels --- often combined with packet interarrival times --- (e.g.,
\cite{Singh:94,bang11:_adapt_clock_recov_mechan_havin}), or timestamps (e.g.,
\cite{noro99:_clock_mpeg,su01,aweya04:_circuit}). Because the interarrival times
and the buffer levels cannot provide the information on the source clock
frequency in case of aperiodic packet streams, only timestamp-based schemes can
be applicable to the asynchronous SCFR through aperiodic packet streams. To the
author's best knowledge, however, SCFR schemes --- including timestamp-based
ones --- have never been applied for and tested with aperiodic packet streams in
the literature.

The jitter time-stamp (JTS) approach \cite{su01} is close to our work in that it
was tested with VBR traffic. It, however, is devised only for on-off type
traffic which generates a periodic stream of fixed-size packets during \emph{on}
periods. Note that the main purpose of JTS is the recovery of \emph{source rate}
defined as the number of packets generated per unit period --- rather than the
source clock frequency itself --- with minimal-length timestamps in the context
of AAL2. For this purpose the JTS generates a clock signal used to deliver
packets to the application layer with active and inactive periods which are
detected from the received on-off type traffic pattern. As for the recovery of
source clock frequency, the JTS does not provide any direct way to estimate
and/or recover it; it suggests an adaptive adjustment of jitter bias, which may
result from changing network conditions and drifting clocks. This procedure,
however, is mainly to eliminate any constant bias in dejittering process but not
for the effects of clock frequency drift.

In this paper we propose two new asynchronous SCFR schemes for general aperiodic
packet streams based on timestamps, which are suitable for the implementation in
real systems, and carry out a comparison study with the conventional
phase-locked loop (PLL)-based scheme.
\section{Clock Frequency Ratio Estimation}
\label{sec-2}
\subsection{Regression through the Origin (RTO) Model}
\label{sec-2-1}

To recover the source clock frequency without a common reference clock or any
reference period in packet streams, we need timestamps in received packets which
delivers the information on departure times of packets. The SCFR through
periodic packet streams is a special case where the constant packet generation
interval, assumed to be known at both the sender and the receiver through
service specifications, can be used to extract this information instead of
timestamps.

\begin{figure}[!tpb]
\begin{center}
\includegraphics[width=.8\linewidth,trim=55 60 40 40,clip=true]{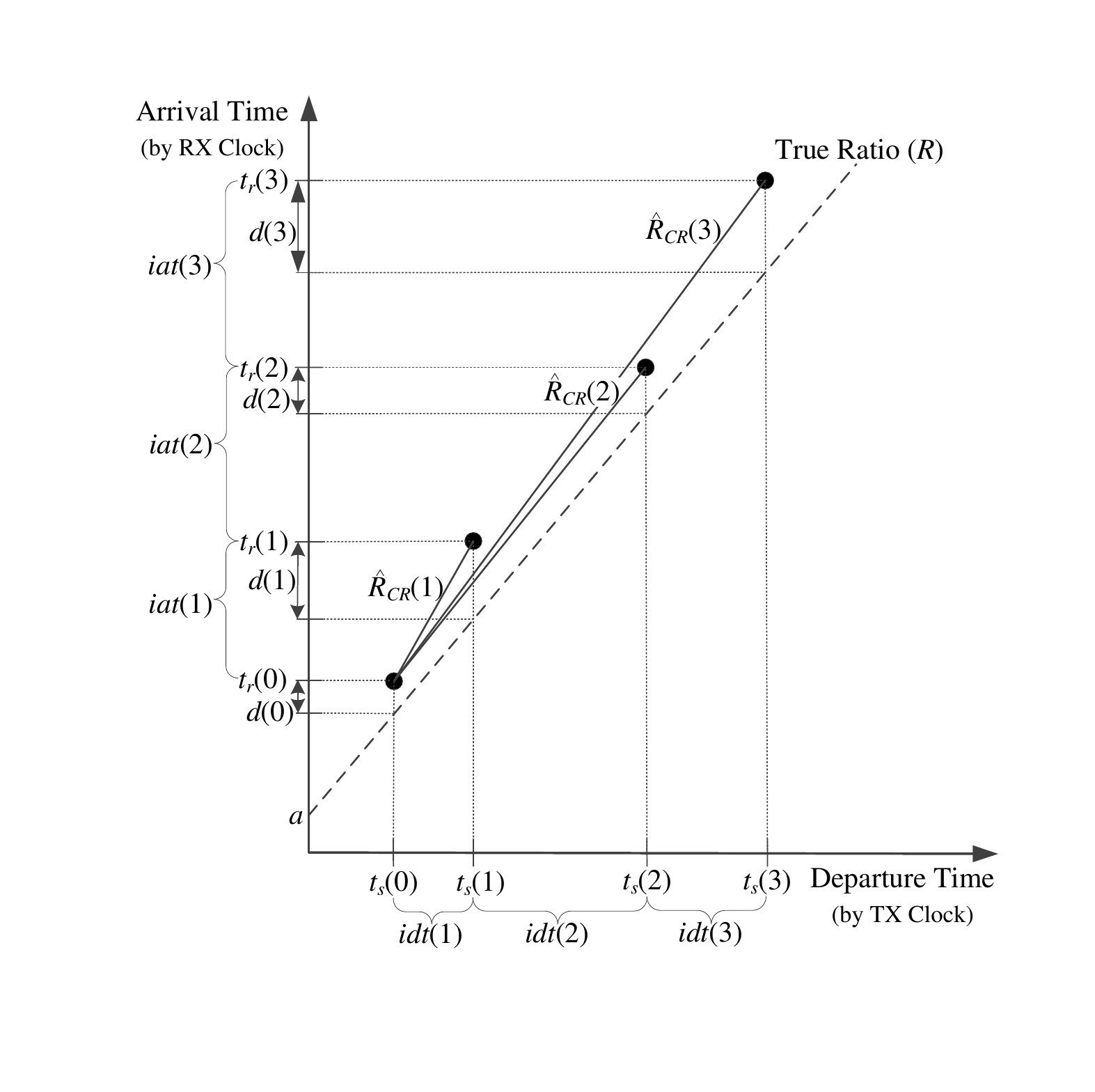}
\end{center}
\caption{Illustration of interarrival times, interdeparture times, and their
cumulative ratios.}
\label{fg:cumulative_ratio}
\end{figure}

Let $t_s(k)$ and $t_r(k)$ are the \(k\)th packet timestamp generated by the
source clock whose frequency is $f_s$ and the corresponding arrival time
measured with a receiver clock whose frequency is $f_r$, respectively. As
illustrated in Fig. \ref{fg:cumulative_ratio}, the arrival time $t_r(k)$
includes a packet delay measured in the receiver clock (i.e., $d(k)$) whose true
value is unknown.

The arrival times, the timestamps, and the packet delays are modeled by linear
regression as follows: For $k \geq 0$,
\begingroup
\setlength{\arraycolsep}{0.0em}
\begin{eqnarray}
\label{eq:linear_regression}
t_r(k) &{=}& R t_s(k) + a + d(k) \nonumber\\
&{=}& R t_s(k) + \alpha + e_k,
\end{eqnarray}
\endgroup
where $R$ is the ratio of the receiver clock frequency to the source clock
frequency, also called \emph{frequency drift}, $a$ is the intercept corresponding to
the initial \emph{time offset} between the clocks,
$\alpha{\triangleq}\left(a{+}\min_{k}d(k)\right)$, and \(e_k{\triangleq}{+}\left(d(k){-}\min_{k}d(k)\right)\). This model of
(\ref{eq:linear_regression}) has been extensively studied in a slightly
different context of trend \cite{trump01:_maxim} or clock skew \cite{Moon:99}
estimation where not only frequency ratio but also phase difference between the
clocks are to be estimated.

Because we need to estimate only the ratio $R$ for SCFR, however, we can
reformulate (\ref{eq:linear_regression}) as a linear \emph{regression through the origin (RTO)} model without a constant term (i.e., $\alpha$)
\cite{eisenhauer03:_regres} by subtracting initial values from arrival times,
timestamps, and packet delays: For $k \geq 1$,
\begin{equation}
\label{eq:linear_rto}
\tilde{t}_r(k) = R \tilde{t}_s(k) + \tilde{d}(k),
\end{equation}
where $\tilde{t}_r(k){\triangleq}t_r(k){-}t_r(0)$,
$\tilde{t}_s(k){\triangleq}t_s(k){-}t_s(0)$, and
$\tilde{d}(k){\triangleq}d(k){-}d(0)$. $\tilde{d}(k)$ represents a noise process
with a zero mean. From Fig. \ref{fg:cumulative_ratio} we can see that
$\tilde{t}_r(k)$ and $\tilde{t}_s(k)$ correspond to the sum of interarrival
times (i.e., $\sum_{i=1}^{k}iat(i)$) and the sum of interdeparture times (i.e.,
$\sum_{i=1}^{k}idt(i)$), respectively. For actual implementation, we use these
representation by interarrival and interdeparture times for better processing of
the wrap-around of counter values.

\begin{figure}[!tpb]
    \centering
    \includegraphics[width=\linewidth,trim=33 25 33 20,clip=true]{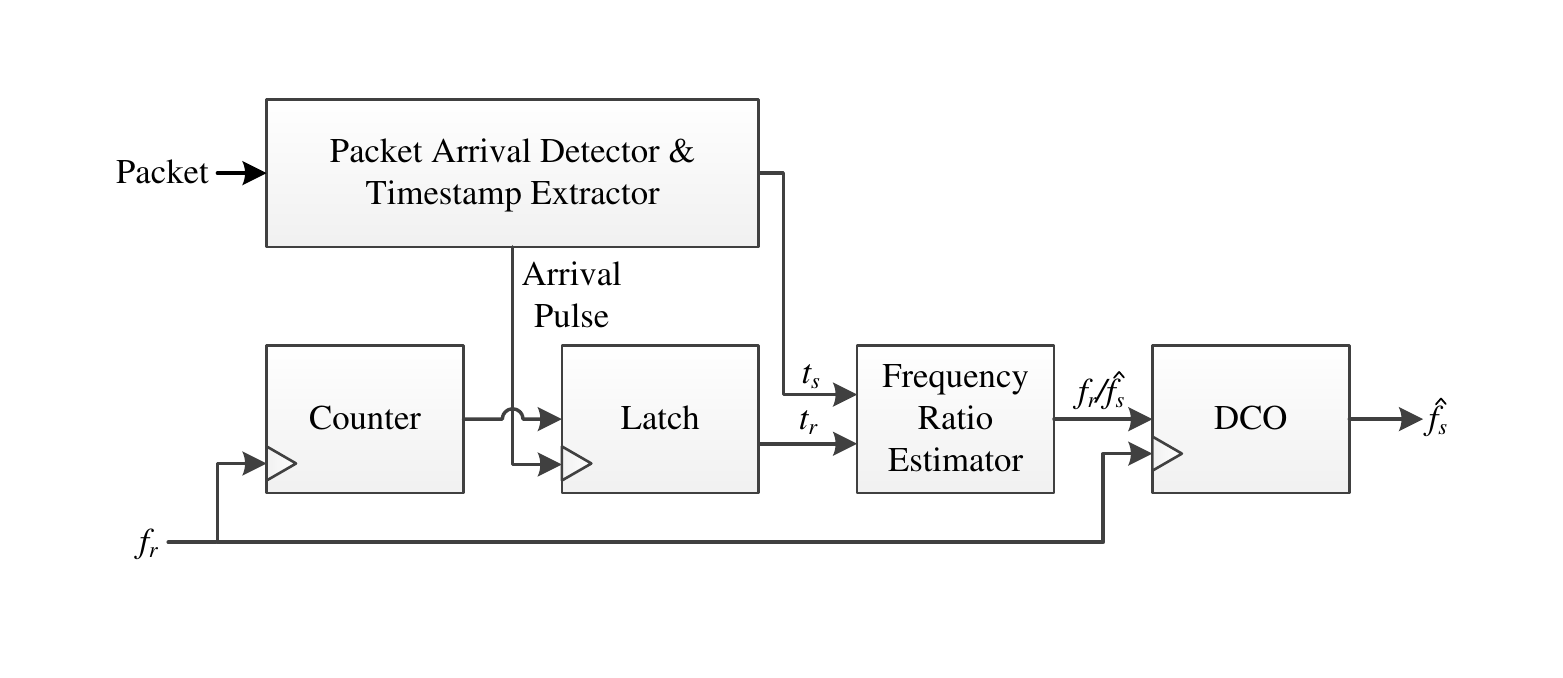}
    \caption{DCO-based implementation of asynchronous SCFR schemes.}
    \label{fg:async_scfr_implementation}
\end{figure}

Fig. \ref{fg:async_scfr_implementation} shows a digitally-controlled oscillator
(DCO)-based implementation of asynchronous SCFR schemes as suggested in
\cite{aweya06:_clock}, where the estimated clock frequency ratio directly drives
a DCO to generate a desired source clock.
\subsection{Least Squares (LS) Estimator}
\label{sec-2-2}

From the linear RTO model of (\ref{eq:linear_rto}), the clock frequency ratio
can be estimated recursively using the recursive least squares (RLS) method
\cite{Ljung:87} as follows: For $k \geq 1$,
\begingroup
\setlength{\arraycolsep}{0.0em}
\begin{eqnarray}
\label{eq:rls_rto}
\hat{R}_{LS}(k) &{=}& \hat{R}_{LS}(k-1) \nonumber\\
&&{+}\:P(k) \tilde{t}_s(k) \left(\tilde{t}_r(k){-}\tilde{t}_s(k)\hat{R}_{LS}(k-1)\right), \\
P(k) &{=}& P(k-1){-}{\displaystyle \frac{P^2(k-1)\tilde{t}_s^2(k)}{1{+}P(k-1)\tilde{t}_s^2(k)}}.
\end{eqnarray}
\endgroup
For initialization of the RLS algorithm, we need to set values for
$\hat{R}_{LS}(0)$ and $P(0)$ from apriori information like nominal values of
clock frequencies \cite{Kim:00-1}. Note that the RLS equations for the linear
RTO model are much simpler than typical linear regression models (e.g., those in
\cite{noro99:_clock_mpeg}) and does not depend on any assumption on the noise
process.
\subsection{Cumulative Ratio (CR) Estimator}
\label{sec-2-3}

If a packet stream is periodic where $idt(k){=}c$ for a constant value $c$, we
can estimate the clock ratio based on the ratio of an interarrival time to an
interdeparture time at each packet arrival like many SCFR schemes based on
packet interarrival times, i.e.,
\begin{equation}
\label{eq:independent_ratio}
\frac{iat(k)}{idt(k)} = R+\frac{d(k)-d(k-1)}{c},~~~~k=1,~2,~3,~\ldots
\end{equation}
In case of an aperiodic packet stream, however, this ratio results in a biased
estimate of the clock frequency ratio because the noise component divided by
varying interdeparture time, i.e,
\begin{equation}
\label{eq:ir_noise}
\frac{d(k)-d(k-1)}{t_s(k)-t_s(k-1)},
\end{equation}
has non-zero average in general due to the dependency between the packet delay
and the interdeparture time, which has been observed during our
simulations.

Fig. \ref{fg:cumulative_ratio} suggests an alternative approach for unbiased
estimation of the clock frequency ratio for aperiodic packet streams through the
ratio of cumulated interarrival times to cumulated interdeparture times. In this
case the clock ratio is recursively estimated at each packet arrival as follows:
For $k \geq 1$,
\begingroup
\setlength{\arraycolsep}{0.0em}
\begin{eqnarray}
\label{eq:cr_iter}
\hat{R}_{CR}(k) &=& \frac{A(k)}{D(k)}, \\
A(k) &=& A(k-1) + iat(k), \\
D(k) &=& D(k-1) + idt(k),
\end{eqnarray}
\endgroup
where $A(0){=}D(0){=}0$. $\hat{R}_{CR}(k)$ can be written as follows:
\begin{equation}
\label{eq:cr}
\hat{R}_{CR}(k) = \frac{\sum_{i=1}^{k}iat(i)}{\sum_{i=1}^{k}idt(i)} = R + \frac{\tilde{d}(k)}{\tilde{t}_s(k)}.
\end{equation}
Note that the noise component becomes zero as time goes to infinity irrespective
of its statistical characteristics. Also note that there are no design parameter
values or initial values to set unlike other SCFR schemes.
\section{Simulation Results}
\label{sec-3}

We carried out a comparison study of the proposed SCFR schemes with the
conventional PLL-based one \cite{noro99:_clock_mpeg,aweya04:_circuit} using an
aperiodic VBR video packet stream from realistic simulation of an access
network, where there are 16 subscribers connected through 100-Mb/s user-network
interfaces (UNIs) to a shared access network with the same line rate of 100
Mb/s, each of which uses hypertext transport protocol (HTTP), file transfer
protocol (FTP) and user datagram protocol (UDP) streaming video traffic with a
common application server in the network with a combined traffic rate of about 2
Mb/s per subscriber. The backbone line rate connecting the access network and
the server and the end-to-end round-trip time are set to 1 Gb/s and 10 ms,
respectively. We use session-level models based on user behaviors for HTTP and
FTP traffic. As for UDP streaming video, we generate traffic using a real trace
for a common intermediate format (CIF) ``Silence of The Lambs'', a VBR-coded
H.264/advanced video coding (AVC) clip from Arizona state university (ASU) video
trace library \cite{Auwera:08-1}; 30 frames are generated per second, and
transmissions of the packets from a frame are either evenly spread over a frame
period (\emph{with frame spreading}) or transmitted at a fixed rate at the beginning
of each frame (\emph{without frame spreading}), the latter resulting in on-off type
traffic. The average and the variance of the resulting packet interdeparture
times in a reference clock (i.e., a global simulation clock) are 5.9026 ms and
0.046159 ms with frame spreading and 5.9051 ms and 0.13472 ms without frame
spreading, respectively. For details of the traffic models and their parameter
values for the simulation, readers are referred to \cite{Kim:12-2}.

The source clock frequency at the server, which is used for generating 32-bit
real-time transport protocol (RTP) timestamps for the streaming video, is set to
90.018 kHz with 200-ppm deviation from the nominal value of 90 kHz per RFC 3551
\cite{rfc:3551}. This clock frequency divided by 50 (i.e., 1.80036 kHz) is also
used to transmit packets during active periods when frame spreading is not used;
this amounts to the rate of about 21 Mb/s with 1460-byte RTP payload, which is
big enough to handle the peak frame rate of the trace (i.e., 19.61616 Mb/s). In
case of LS and CR estimators, arrival times are measured using a 48-bit counter
driven by the receiver clock with frequency of 15.9968 MHz. For the PLL-based
scheme, the free-running frequency of the voltage-controlled oscillator (VCO) is
set to 89.982 kHz, and, instead of a counter, floating-point numbers are used to
represent arrival times.\footnote{The quantization error in phase difference calculation due to the use of
a counter makes the PLL never converge to its steady state during the simulation
with aperiodic packet streams; this does not happen with periodic packet
streams. } The local clock frequencies for both the cases
amount to the frequency deviation of 200 ppm from the nominal values of 16 MHz
and 90 kHz, respectively.\footnote{The local clock frequency should be high enough (i.e., $f_r \geq 50 f_s$)
in order to reduce quantization errors related with the DCO-based clock
generation \cite{aweya04:_circuit}. This is not the case for VCO-based clock
generation. }

For the LS estimator, we set the initial values of $\hat{R}(0)$ and $P(0)$ to
the ratio of nominal clock frequencies (i.e., 16000000/90000) and 10,
respectively. For the PLL-based scheme, we assume a critical damping (i.e.,
damping factor of 0.707) and set the gains of proportional and integral paths to
0.0001 and 0.000001, respectively. Note that the CR estimator has no parameter
values to set.

\begin{figure}[!tpb]
\centering
\includegraphics[width=0.85\linewidth,trim=15 10 15 20,clip=true]{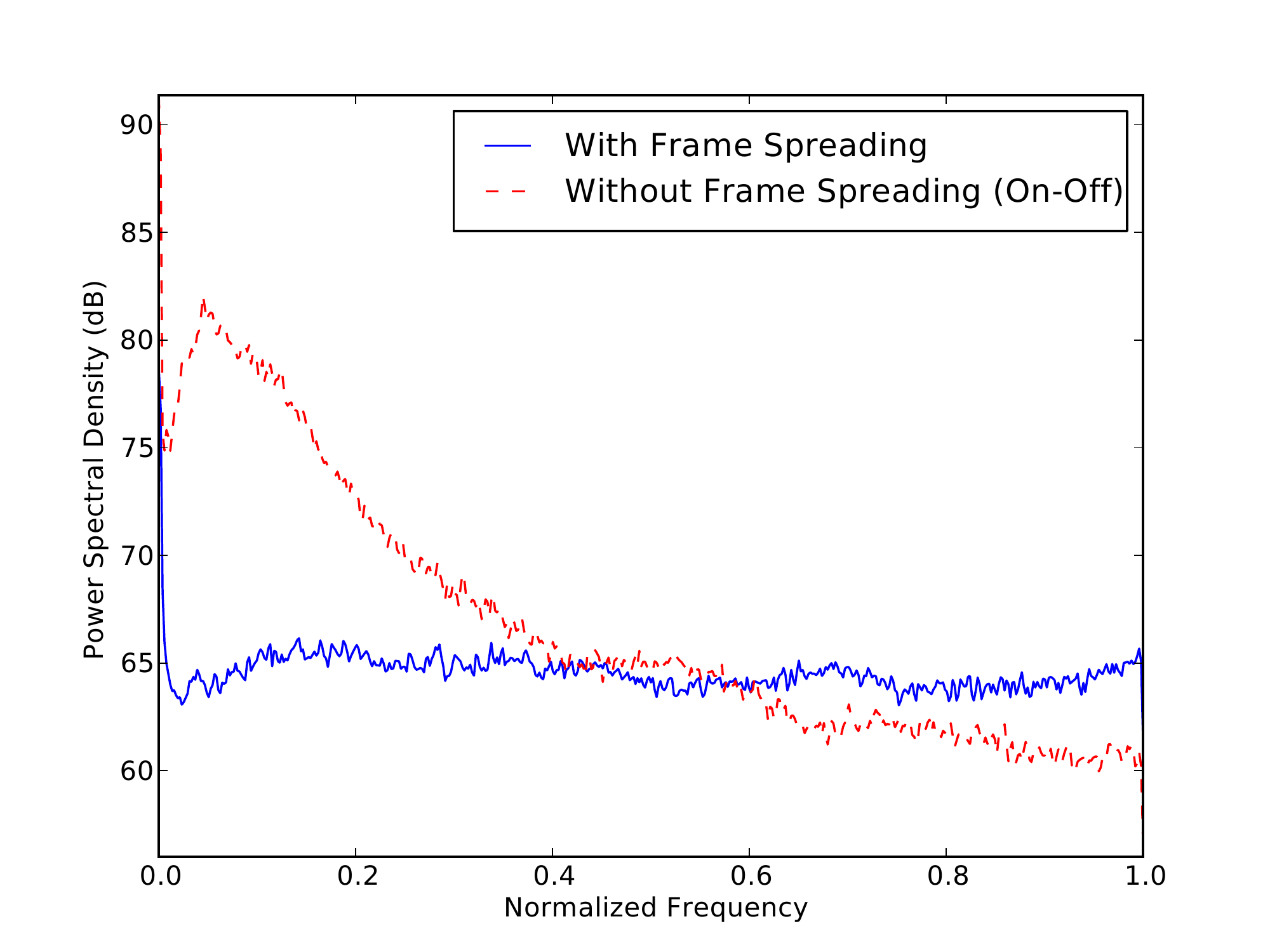}\\
{\scriptsize (a)} \\
\includegraphics[width=0.85\linewidth,trim=15 10 15 20,clip=true]{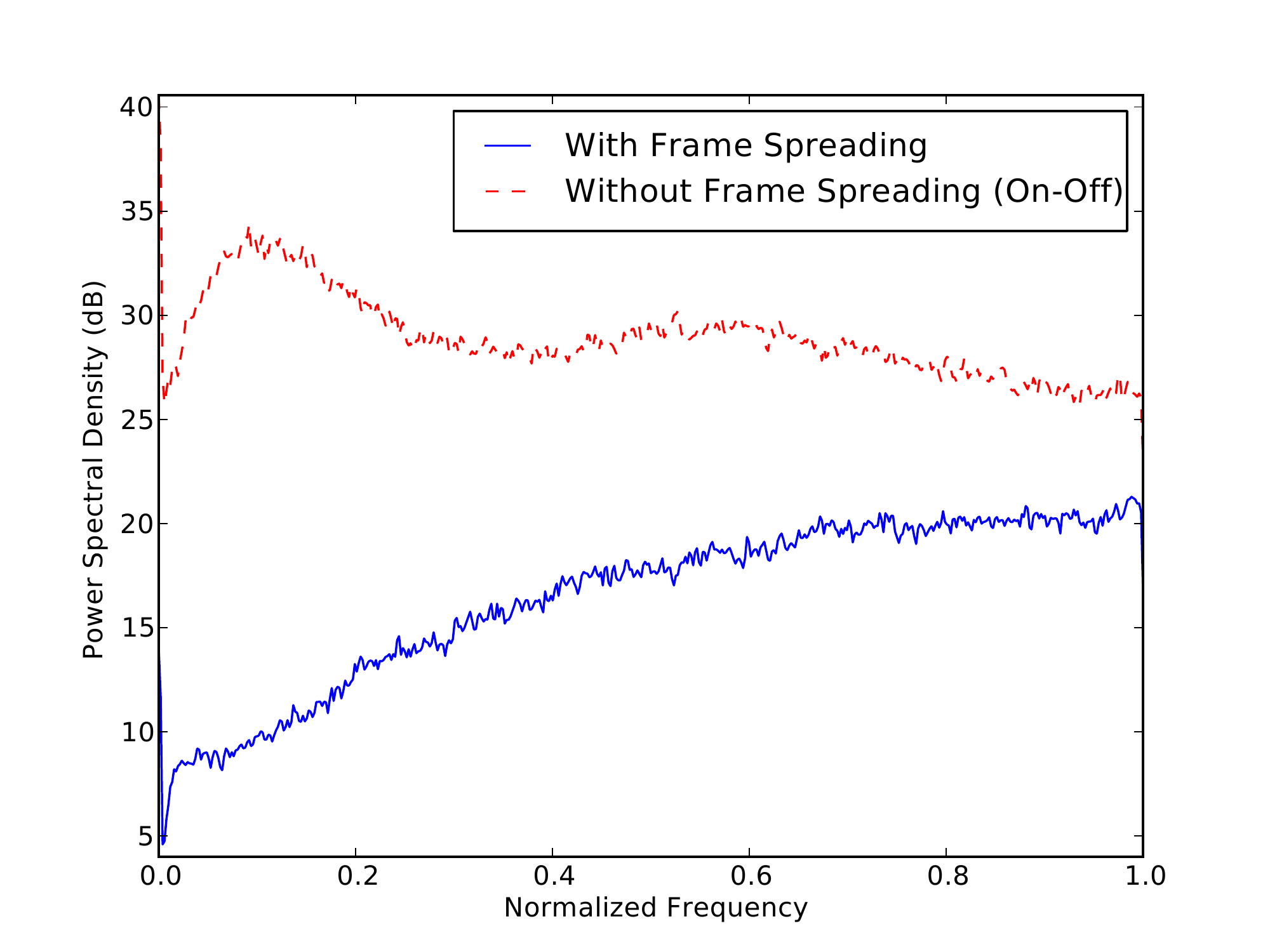}\\
{\scriptsize (b)}
\caption{Power spectral density of the noise component in SCFR for UDP streaming video:
(a) $\tilde{d}(k)$ in (\ref{eq:linear_rto}) and (\ref{eq:cr}); (b)
$\frac{d(k)-d(k-1)}{t_s(k)-t_s(k-1)}$ in (\ref{eq:ir_noise}).}
\label{fg:N16_n1_vlow_noise_psd}
\end{figure}

Fig. \ref{fg:N16_n1_vlow_noise_psd} shows the power spectral densities of the
noise components from the simulation, i.e., $\tilde{d}(k)$ in
(\ref{eq:linear_rto}) and (\ref{eq:cr}) for the proposed SCFR schemes and
$\frac{d(k)-d(k-1)}{t_s(k)-t_s(k-1)}$ in (\ref{eq:ir_noise}) for the PLL-based
scheme, which clearly illustrate the effect of frame spreading and resulting
traffic pattern: We observe that the use of frame spreading lowers the level of
low-frequency components and flattens the overall spectrum, more evidently in
the case of $\tilde{d}(k)$ as shown in Fig. \ref{fg:N16_n1_vlow_noise_psd} (a).

\begin{figure}[!tpb]
\centering
\includegraphics[width=0.85\linewidth,trim=10 15 10 20,clip=true]{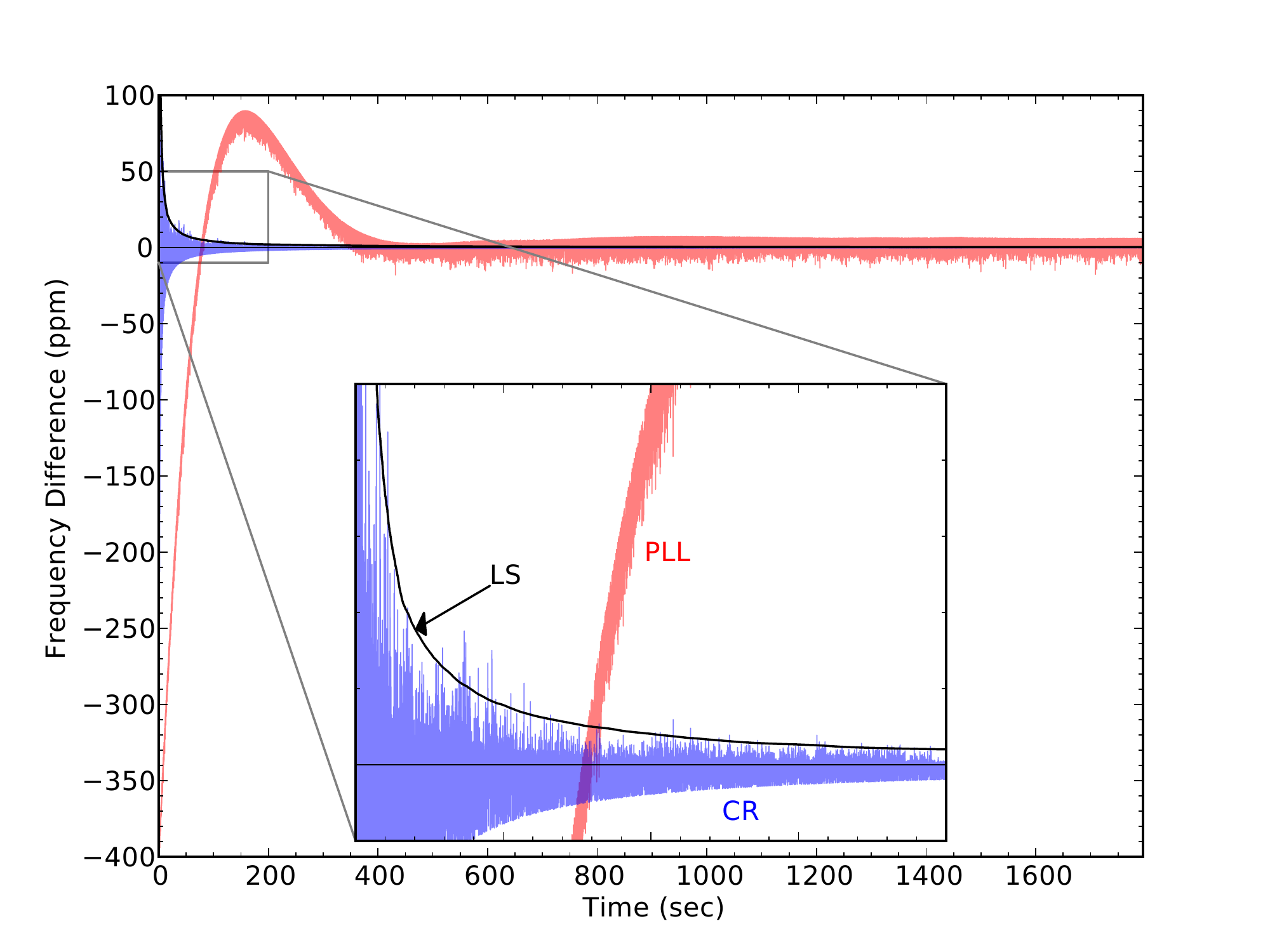} \\
{\scriptsize (a)} \\
\includegraphics[width=0.85\linewidth,trim=10 15 10 20,clip=true]{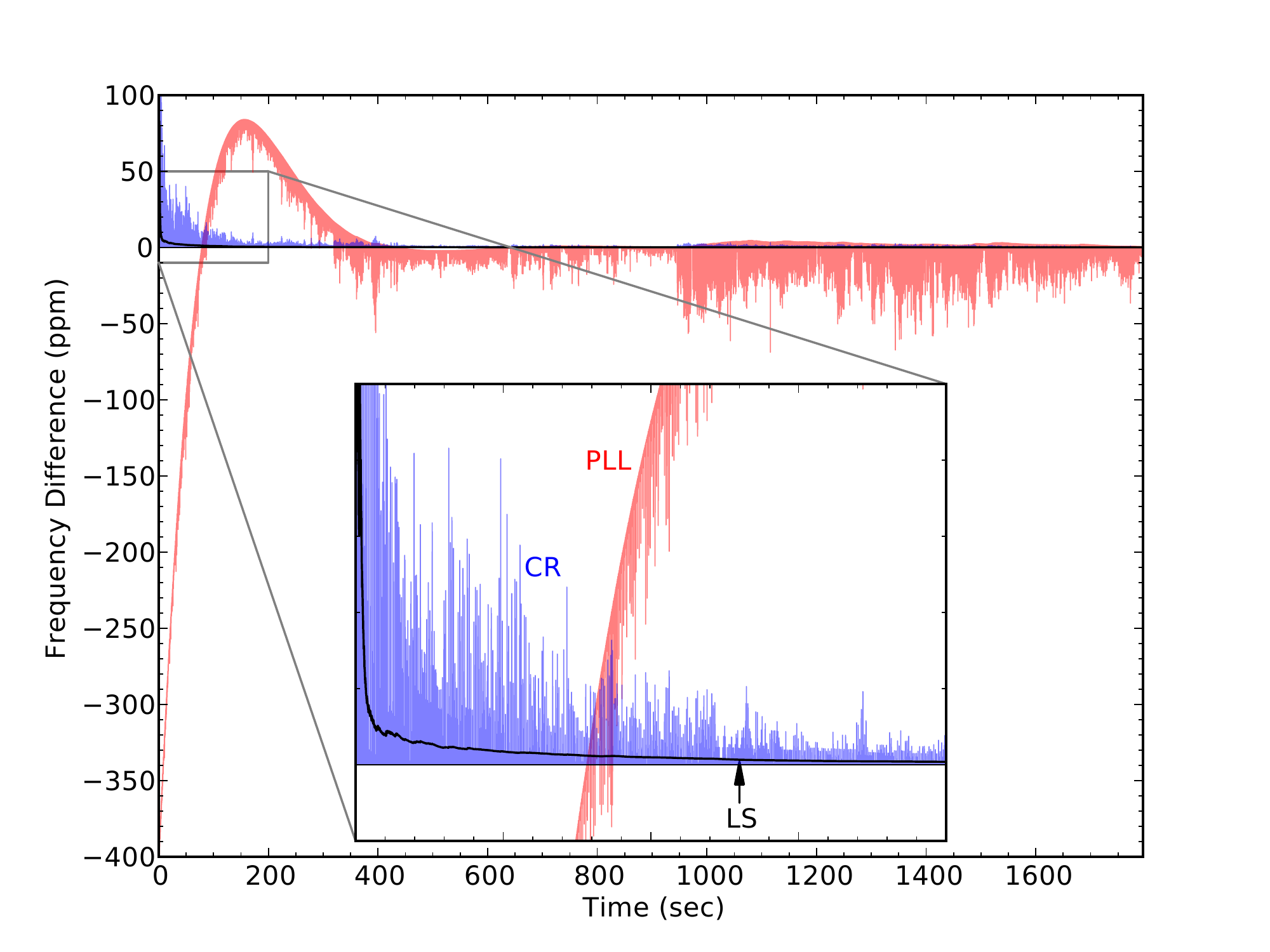} \\
{\scriptsize (b)}
\caption{Frequency estimation errors for UDP streaming video: (a) With frame spreading;
(b) without frame spreading (i.e., on-off traffic).}
\label{fg:N16_n1_vlow_fd}
\end{figure}

Fig. \ref{fg:N16_n1_vlow_fd} shows the frequency estimation errors from the
three asynchronous SCFR schemes for the UDP streaming video.
It is clear from the results that both the proposed schemes outperform the
PLL-based one in terms of the speed of convergence and the size of residual
errors. Of the proposed schemes, the CR provides nearly equivalent performances
with some fluctuations, which is remarkable considering its very low
computational complexity (i.e., two storages for $A(k)$ and $D(k)$ and one
division and two additions per iteration). Note that the PLL-based scheme is
very sensitive to the choice of parameter values, especially the VCO gain; the
results shown in Fig. \ref{fg:N16_n1_vlow_fd} are from the parameter values
which produce the best results after many trials.

Comparing the results with and without frame spreading, we found that the frame
spreading reduces fluctuations in both the CR estimator and the PLL-based
scheme, while the on-off traffic pattern resulting from no frame spreading
improves the convergence speed of the LS estimator. As shown in
Fig. \ref{fg:N16_n1_vlow_noise_psd}, the frame spreading significantly changes
the power spectral densities of noise components, but the results in
Fig. \ref{fg:N16_n1_vlow_fd} show that the proposed schemes are less susceptible
to the change in noise component statistics than the PLL-based one.
\section{Conclusions}
\label{sec-4}

Aperiodic packet streams in asynchronous networks brings major challenges to the
SCFR; one of them is distortion of packet jitter statistics as discussed with
(\ref{eq:ir_noise}), which leads into biased estimation of the clock frequency
ratio between the source and the receiver clocks. In this letter we have
proposed two asynchronous SCFR schemes for aperiodic packet streams, one based
on the RLS method and the other based on simple heuristics of cumulative ratio
of interarrival and interdeparture times. Both the schemes are not based on any
assumption on the noise process and greatly reduce computational complexity
through the formulation of SCFR with timestamps as a linear RTO
model. Simulation results demonstrate that the proposed schemes outperform the
conventional PLL-based one in terms of the speed of convergence and the size of
residual errors in estimating the source clock frequency.

\bibliographystyle{IEEEtran}
\bibliography{IEEEabrv,kks}

\end{document}